\documentclass[aip, superscriptaddress, amsmath,amssymb,reprint]{revtex4-2}
\usepackage{graphicx}
\usepackage{dcolumn}
\usepackage{bm}
\usepackage[utf8]{inputenc}
\usepackage[T1]{fontenc}
\usepackage{mathptmx}
\usepackage{etoolbox}
\usepackage{xcolor}

\makeatletter
\def\@email#1#2{%
 \endgroup
 \patchcmd{\titleblock@produce}
  {\frontmatter@RRAPformat}
  {\frontmatter@RRAPformat{\produce@RRAP{*#1\href{mailto:#2}{#2}}}\frontmatter@RRAPformat}
  {}{}
}%
\makeatother
\begin{document}

\preprint{AIP/123-QED}

\title{Model-free prediction of multistability using echo state network}
\author{Mousumi Roy}
\affiliation{Department of Physics, Central University of Rajasthan, Ajmer, Rajasthan, India - 305817}
\author{Swarnendu Mandal}%
\affiliation{Department of Physics, Central University of Rajasthan, Ajmer, Rajasthan, India - 305817}
\author{Chittaranjan Hens}
\affiliation{Center for Computational Natural Sciences and Bioinformatics, International Institute of Information Technology, Gachibowli, Hyderabad, India - 500 032}

\author{Awadhesh Prasad}
\affiliation{Department of Physics \& Astrophysics, University of Delhi, Delhi, India - 110 007}

\author{ N.V. Kuznetsov}
\affiliation{Department of Applied cybernetics, Saint Petersburg University, St. Petersburg, Russia}
\affiliation{Department of Mathematical Information Technology, University of Jyvaskyla, Jyvaskyla, Finland}
\affiliation{Institute for Problems in Mechanical Engineering of the Russian Academy of Sciences, St. Petersburg, Russia}

\author{Manish Dev Shrimali}
\affiliation{Department of Physics, Central University of Rajasthan, Ajmer, Rajasthan, India - 305817}

\date{\today}

\newcommand{\clb}{\color{blue}}
\newcommand{\clr}{\color{red}}

\begin{abstract}
In the field of complex dynamics, multistable attractors have been gaining a significant attention due to its unpredictability in occurrence and extreme sensitivity to initial conditions. Co-existing attractors are abundant in diverse systems ranging from climate to finance, ecological to social systems. In this article, we investigate a data-driven approach to infer different dynamics of a multistable system using echo state network (ESN). We start with a parameter-aware reservoir and predict diverse dynamics for different parameter values. Interestingly, machine is able to reproduce the dynamics almost perfectly even at distant parameters which lie considerably far from the parameter values related to the training dynamics. In continuation, we can predict whole bifurcation diagram significant accuracy as well.
We extend this study for exploring various dynamics of multistable attractors at unknown parameter value. While, we train the machine with the dynamics of only one attarctor at parameter $p$, it can capture the dynamics of co-existing attractor at a new parameter value $p+\Delta p$.
Continuing the simulation for multiple set of initial conditions, we can identify the basins for different attractors. We generalize the results by applying the scheme on two distinct multistable systems.    
\end{abstract}

\maketitle

\begin{quotation}
The occurrence of multistability is evident to exist in many daily life complex dynamical systems like climate, financial market, ecological systems and many more. Often the unpredictable behavior of such systems lead to undesired scenario. Hence, it is important to analyse the dynamical behaviors of these systems. But often we do not have the exact mathematical model to mimic the behavior of many such systems. In this study, we use a data-driven machine learning approach to infer the dynamics of multistable systems. We discuss that, a few time series data with its corresponding system parameters are enough to capture essence of complex dynamics for a broad parameter range.
Moreover, training the system for a given parameter, it is possible to identify the presence of unknown co-existing attractors for a different parameter even with their basin of attraction.

\end{quotation}

\section{\label{sec:introduction}Introduction}
In the field of non-linear dynamics, multistable attractors possess different stable dynamics\cite{feudel2008complex,tanaka2011multistable,hens2015extreme,hens2012obtain} at a fixed parameter set solely depending on the initial conditions. Its dynamical uncertainty and extreme dependency on the initial conditions make it difficult to acquire knowledge of the overall system dynamics. Additionally, no prior knowledge of the basins corresponding to different stable dynamics makes it harder while working with such systems. Basin can be smooth or riddled-like~\cite{chaudhuri2014complicated,sharma2015controlling}. In reality, multistable attractors are present in diverse systems, such as, ecology~\cite{feudel2018multistability,suzuki2021energy}, electrical machines~\cite{louodop2019extreme}, financial~\cite{cavalli2017real}, laser\cite{masoller1994coexistence} to name a few. Because of its presence in various fields, the nature of its dynamics, as well as the basins of such multistable attractors, are significantly important to be identified. Most importantly, as in most of these cases, the exact mathematical form of the system is unknown. Since last few years, a large number of theoretical and numerical studies have been making a great effort to shed light on this dimension. However, most of the studies are focused on the route of controlling multistability within the system~\cite{pisarchik2014control,sharma2015controlling,shajan2022controlling}. Despite these enormous attempts by researchers, any straightforward method for model-free prediction of different dynamics and basins is still rare. \par
Motivated from this backdrop, in this study we address this issue using a model-independent machine learning technique. Here, we use a simple yet extremely powerful tool Echo State Network (ESN)~\cite{jaeger2001echo}, which is popular for its training cost efficiency and architectural simplicity. Based on the basic mechanism, various form of ESN is available targeting to make a balance between efficiency and simplicity of the machine architecture~\cite{cui2012architecture,ma2016functional,gauthier2021next,pathak2018model}. Because of its uncomplicated application procedure, ESN appears as one of the effective tools in diverse areas, ranging from neuroscience~\cite{tanaka2019recent,kim2019decoding,hramov2021physical} to robotics~\cite{hauser2021physical}, speech recognition~\cite{skowronski2007automatic} to language processing~\cite{skowronski2007noise}, network connectivity inference~\cite{banerjee2021machine,banerjee2022inference} to epidemic spreading~\cite{ghosh2021reservoir}, etc. In this context, ESN also takes a significant role in predicting diverse dynamical characteristics, such as, chaotic time series prediction~\cite{weng2019synchronization,borra2020effective,carroll2018using}, Lyapunov exponents~\cite{pathak2017using}, cluster synchronization~\cite{saha2020predicting}, burst synchronization~\cite{roy2022role}, spatio-temporal dynamics~\cite{zimmermann2018observing}, etc. Another significant direction is to explore the way of optimizing training procedures and choosing hyperparameters~\cite{lukovsevivcius2012practical} of ESN. Despite these, researchers use several physical systems as a reservoir as well, developing an intriguing field of physical reservoir\cite{tanaka2019recent,mandal2022machine}.    \par 
Some recent studies have shown the identification of sudden dynamical changes using ESN depending on system parameters. For example, in a coupled system, ESN can predict the critical coupling strengths at which the first order transition occurs~\cite{fan2021anticipating}. Furthermore, Xiao et al. have shown the prediction of amplitude death due to a parameter drift~\cite{xiao2021predicting}. Recently, another interesting study provided a different scheme for training, where they make scaling of the system parameter and get diverse dynamics by switching the parameter at different values, which leads to identifying period doubling route towards chaos~\cite{kim2021teaching}. A recent study has also explored different attractors of multistable attractors using a machine learning architecture~\cite{rohm2021model}. However, the prediction accuracy can be improved. \par

Based on the previous investigations and to fill the gap in the field of multistability, we exploit a data-driven machine learning approach to locate the basins and diverse dynamics of the systems exhibiting more than one stable attractor at a particular parameter set. We proceed for this study in two directions using ESN: The first one is the identification of various dynamics using only a few system parameters and corresponding time series for training. To be specific, we train the machine with three time series data along with their respective parameter values to build a parameter-aware reservoir\cite{xiao2021predicting} and it can predict the dynamics for a range of parameter with great accuracy. In contrast to an earlier attempt\cite{xiao2021predicting}, here we show the machine's ability to infer the dynamics at distant parameters  from the training data even when the training dynamics are less complex, for example, mere periodic. The second dimension is to further extend this study with an attempt to identify different dynamics and the corresponding basins of multistable attractors.  Unlike a previous attempt\cite{rohm2021model}, in our study, we use a different architecture that significantly improves the level of accuracy in prediction. Furthermore, the natural unpredictability and initial condition dependency of multistable systems in real life make it quite difficult to deal with those systems without having any prior knowledge. Inspired by this backdrop, in this study we use a model-free data-driven technique to classify different attractors as well as their basins of a multistable system. We train the ESN using the dynamics of only one attractor and predict the others' dynamics, and basins for every attractor as well, for an unknown system parameter value.\par
We arrange the article as follows: in Sec.~\ref{sec:esn}, we elaborately discuss the basic architecture of ESN that we use for this investigation. Sec.~\ref{sec:result} is devoted to the results. This section starts with a description of the original dynamics of a system that we use for the study. Next, we arrange all the results such as the prediction of different dynamics corresponding to different parameter values, and obtain the bifurcation diagram. Furthermore, we identify the basin of multistable attractors depending on a parameter drift. Furthermore, we verify all the results with another model exhibiting bistability between a stable oscillator and a stable fixed point. Finally, we conclude this study with sec.~\ref{sec:conclusion}.

\section{ESN architecture}\label{sec:esn}
The reservoir dynamics is described with the following map as follows:
\begin{equation}\label{eq:esn} 
    \bm{r}(t+1)=(1-a)\bm{r}(t)+a~ \mathrm{tanh}[\mathcal{W}_{\rm{res}}.\bm{r}(t)+\mathcal{W}_{\rm{in}}.u(t)],
\end{equation}
where, $\bm{r}(t)$ represents the state of the reservoir at time $t$. $N_{\rm{res}}$ is the dimension of the reservoir. $\mathcal{W}_{\rm{res}}$ is a random sparse matrix carrying the connection weights among $N_{\rm{res}}$ number of nodes, having spectral radius $\rho$. The leaking parameter $a$ can take values from 0 to 1. It controls the trade-off between the information of reservoir dynamics and non-linear functional value of the input data. We consider an $m$-dimensional time series data for training, therefore, $u(t)$ is the $m+1$-dimensional input data along with the corresponding system parameter value. To incorporate the information of the parameter through training data we take $u(t)$ as the dynamics along with the corresponding parameter value, where, the first $m$ elements are the dynamics of the input signal $\hat{u}(t)$ at time $t$ and the last element is the corresponding parameter, i.e. $$u(t)=[\hat{u}(t)~ p]^T$$.
$\mathcal{W}_{\rm{in}}$ is the matrix that makes connections within the input data and the reservoir. We randomly choose the elements of $\mathcal{W}_{\rm{in}}$ from the uniformly distributed interval $[-\sigma,\sigma]$. We form $W_{\rm{in}}$ in such a way\cite{xiao2021predicting}, that the information of the input data would be equally distributed to the nodes of the reservoir, that means the information of each dimension of input goes to $\frac{m}{N_{\rm{res}}}$ number of reservoir nodes. However, the information of the parameter is connected to all the nodes of reservoir. That makes the reservoir aware about the relation between the dynamics and the respective parameter. \par
We train the machine for $N_p$ different parameter values one by one and successively store the values of $$\bm{r}(t)=[r_1(t), r_2(t), r_3(t),...,r_{N_{\rm{res}}}(t)]$$ following the conventional way\cite{pathak2018model}: 
\begin{equation}
\bm{\tilde{r}}(t)=
    \begin{cases}
        r_n(t) & \text{if } \hspace{0.2cm}n \hspace{0.2cm}\text{is odd}\\
        r^2_n(t) & \text{if } \hspace{0.2cm}n \hspace{0.2cm}\text{is even}
    \end{cases} 
\end{equation}
We store the values of $\bm{\tilde{r}}(t)$ within the matrix $\mathcal{R}$ after ignoring $t_{\rm{trans}}$ time points as the transient dynamics of reservoir. Therefore, $\mathcal{R}$ is a matrix with dimension $N_{res}\times N_p(t_f-t_{\rm{trans}})$  for a training data of length $t_f$. After calculating $\mathcal{R}$, we can get the readout weight matrix from the system of linear equation,
\begin{equation}
\mathcal{U}=W_{\rm{out}}\mathcal{R},
\end{equation}
where, $\mathcal{U}$ is the matrix having targeted data sets with dimension $m\times N_p(t_f-t_{\rm{trans}})$. For this case, the target output is the state of the system at next time step. Now, in order to minimize the error between $\mathcal{U}$ and $\mathcal{R}$, we apply Ridge regression and obtain the read-out matrix $\mathcal{W}_{\rm{out}}$ in the following way,
\begin{equation}\label{eq:ridge_regression}
\mathcal{W}_{\rm{out}}=\mathcal{U}\mathcal{R}^T(\mathcal{R}\mathcal{R}^T+b I)^{-1}    
\end{equation}
where $b$ is the regularization parameter used to avoid the issue of over-fitting. After completing a successful training, machine is now ready to make a prediction depending only on the parameter values.\par
In the prediction process, we replace only the parameter value with the new one and ESN produces the corresponding time series in an auto-iterative way only on the basis of $\mathcal{W}_{\rm{out}}$ and the new parameter value using the following equations:

\begin{equation}\label{eq:prediction1} 
    \bm{r}(t+1)=(1-a)\bm{r}(t)+a~ \mathrm{tanh}[\mathcal{W}_{\rm{res}}.\bm{r}(t)+\mathcal{W}_{\rm{in}}.v(t)],\\
\end{equation}
\begin{equation}\label{eq:prediction2}
v(t)=[\hat{v}(t)~ p_{\rm{new}}]^T
\end{equation}
\begin{equation}\label{eq:prediction3}
\hat{v}(t)=W_{\rm{out}}\bm{\tilde{r}}(t),
\end{equation}
where
\begin{equation}\label{eq:prediction4}
\bm{\tilde{r}}(t)=
    \begin{cases}
        r_n(t) & \text{if } \hspace{0.2cm}n \hspace{0.2cm}\text{is odd}\\
        r^2_n(t) & \text{if } \hspace{0.2cm}n \hspace{0.2cm}\text{is even}
    \end{cases} 
\end{equation}
Repeating Eqs.~\eqref{eq:prediction1}-\eqref{eq:prediction4}, we obtain the machine generated predicted time series at the new parameter values. We start the prediction process from the initial condition where the system possesses a bounded solution. One trivial choice is to start from one of the points lying on the trajectory of the training data. Before starting the generation of predicted time series, the reservoir needs to warm-up for some iterations at new parameter space to get rid of the effect of reservoir state initialization. To do this, we repeat Eqs.~\eqref{eq:prediction1}-\eqref{eq:prediction3} for $w_a$ time points and let the machine produce the predicted time series $\hat{v}(t)$ following the above discussed method.\par
This scheme involves certain hyperparameters, namely spectral radius $(\rho)$, leaking parameter $(a)$, range of the non-zeros values of $W_{in}$ matrix $(\sigma)$ and regularization parameter $(b)$. We optimize these parameters before running predictions for different tasks following a simultaneous optimization technique\cite{griffith2019forecasting,yperman2016bayesian}. We use the root-mean-square error (RMSE) of predicted time series of length $N_p$ with respect to its actual target, averaged over sufficient number of reservoir realizations, as the loss function to be minimized for the optimization process.\par
After obtaining a set of optimal hyperparameters, we build the final machine using those hyperparameters and now machine is ready to predict different dynamics depending on the parameter values and initial conditions. 

\section{Results}\label{sec:result}
To illustrate the previously mentioned reservoir-computing scheme, in this section we consider two models having multistable attractors. For both the models, we train the machine with the dynamics of only one attractor and try to predict the dynamics depending on diverse parameter values and initial conditions. We always choose oscillatory dynamics over the steady-state from the bi-stable system for training to make the reservoir well-informed about the dynamical information of the original system.   
\subsection{\label{sec:original_model}Multistable system}
\begin{figure}
    \centering
    \includegraphics[width=0.5\textwidth]{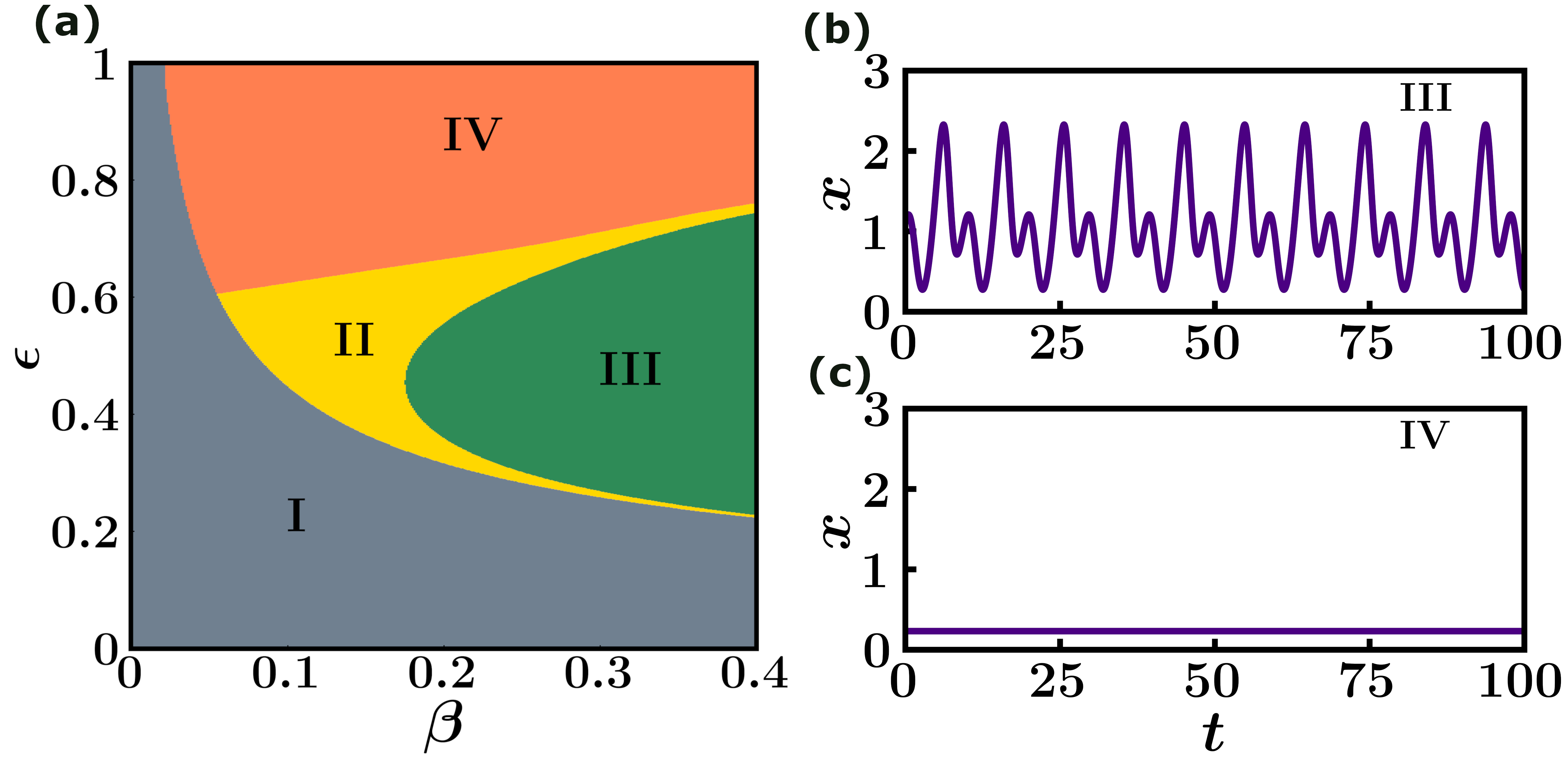}
    \caption{(a) Overall dynamics of the system~\eqref{eq:original_model1} is presented over $(\beta-\epsilon)$ parameter space. Region $\rm{I}$ shows the region of unbounded solution i.e. the system blows up towards infinity. (b) In region $\rm{III}$, the system possesses an unstable fixed point and a stable oscillatory dynamics can be observed. (c) In region $\rm{IV}$ the fixed point becomes stable. A bistability appears between two stable attractors: steady state and oscillatory dynamics in region $\rm{II}$.}  
    \label{fig:original_model1}
\end{figure}

We consider a non-linear system~\cite{sharma2015controlling} having multistability as follows:
\begin{equation}\label{eq:original_model1}
\begin{aligned}
\dot{x} &=y\\
\dot{y} &=x\\
\dot{z} &=-y+3y^2-x^2-xz+\alpha+\epsilon u\\
\dot{u} &=-ku-\epsilon(z-\beta)
\end{aligned}
\end{equation}
Depending on the parameters $\alpha$, $\beta$ and $\epsilon$, system~\eqref{eq:original_model1} exhibits several dynamics. For $\alpha=-0.02$, overall dynamics of the system is shown on a $(\beta-\epsilon)$ parameter space (see Fig.\ref{fig:original_model1}(a). Here, depending on the values of $\epsilon$ and $\beta$, the system does not possess any fixed point in region I, highlighted with grey color, and in the other regions (II, III, IV) it generates a fixed point as follows, 
\begin{equation}\label{eq:fixed_point}
    x^*=\pm\sqrt{\alpha+\frac{\epsilon^2\beta}{k}},y^*=0,z^*=0,u^*=\frac{\epsilon\beta}{k}
\end{equation}
Clearly, on satisfying the condition $(\alpha+\frac{\epsilon^2\beta}{k})>0$ i.e.$\alpha>-\frac{\epsilon^2\beta}{k}$, the system has a real fixed point. Region II (marked by yellow), III (marked by green) and IV (marked by orange) are classified based on the stability of the fixed point. In the region III, the fixed point is unstable produces an oscillatory dynamics and region IV depicts the parameter space of stable steady state. Furthermore, in region II, system has multistability with unbounded state, stable steady state and stable oscillatory dynamics. In this continuation, in Fig.\ref{fig:original_model1}(b), we plot the oscillatory dynamics for region (III) and and in Fig.\ref{fig:original_model1}(c), the steady state dynamics is shown as in region (IV). \par

\subsection{Prediction of dynamics at various parameters}
\begin{figure}
    \centering
    \includegraphics[width=0.5\textwidth]{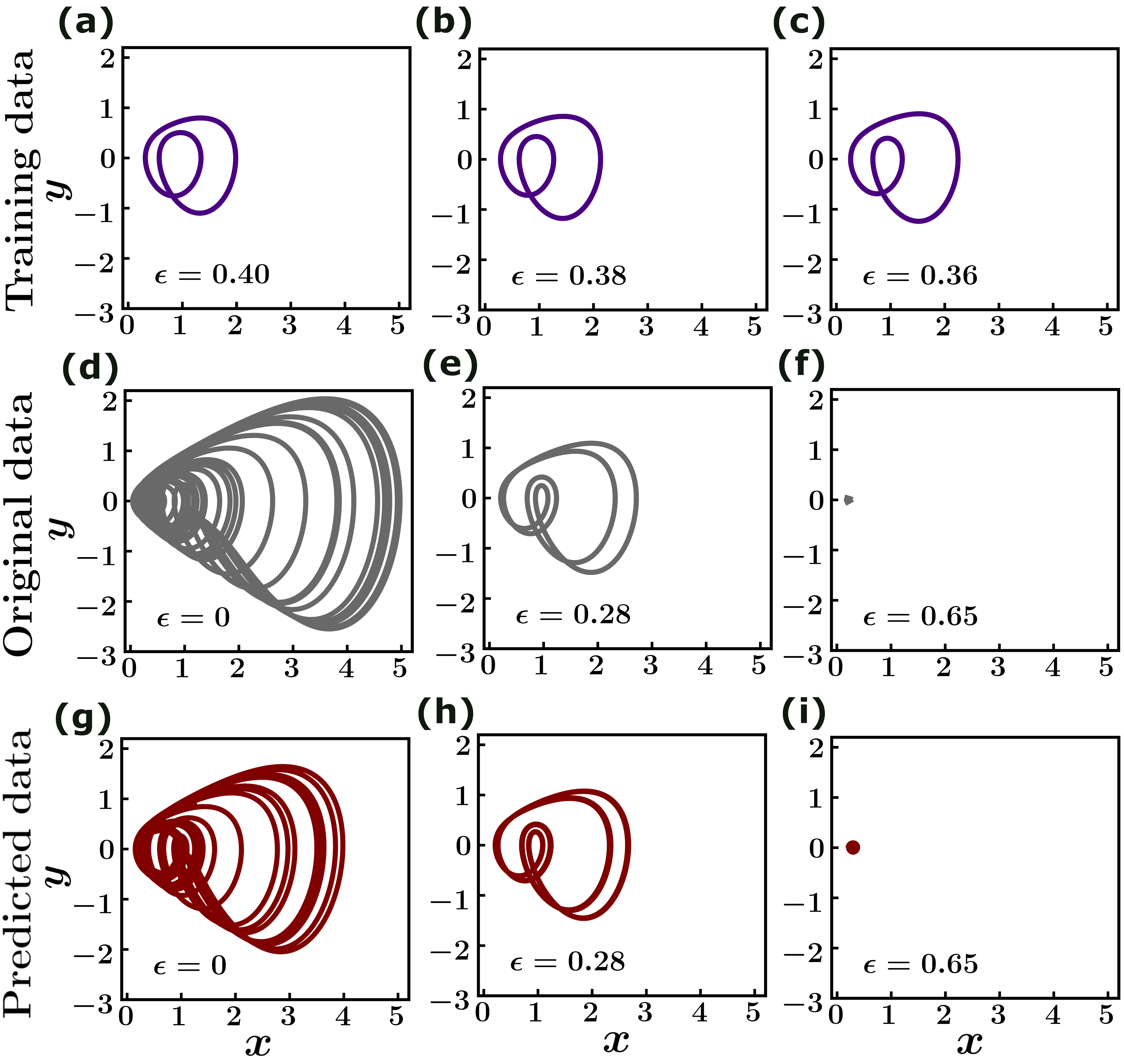}
    \caption{(a)-(c) The training dynamics which are used to train ESN. We use periodic dynamics for $\epsilon=0.4, 0.38, 0.36$. (d)-(f) The original dynamics those we are trying predict using ESN for diverse values of $\epsilon$. (g)-(i) Corresponding predicted dynamics.}  
    \label{fig:predicted_dynamics}
\end{figure}
For this study, we fix the parameter at $\beta=0.15$ and choose $\epsilon$ as the varying parameter. Here, we train ESN with the dynamics at $\epsilon=0.4,~0.38$ and $0.36$. At each parameter value, the system generates periodic oscillation as described in Figs.\ref{fig:predicted_dynamics}(a), (b), (c), and our initial target is to predict different dynamics of the system for completely new parameter values which the machine is unaware of. \par
For numerical simulations, we solve system~\eqref{eq:original_model1} using RK45 numerical scheme with $0.01$ time-step at the values of chosen training parameters. After avoiding sufficient transient dynamics, we use the dynamics of next $N_t=60000$ time points for each of the parametric values to train ESN. We use the dynamics along with the parameter in Eq.~\eqref{eq:esn} and follow the procedure as described in Sec.~\ref{sec:esn}. Now, we avoid first 100 time points as the transients of machine and store all the reservoir states within the the matrix $\mathcal{R}$. Next using Ridge regression (as described in Eq.~\eqref{eq:ridge_regression}), we obtain the readout datasets, and finally the machine is ready to predict the dynamics at new parameter values. We fix the machine hyperparameters at $\rho=0.06, ~\sigma=0.5057,~a=0.385,~b=5\times 10^{-8}$. The reservoir size and number of average random connections within the reservoir are fixed at $N=1200$ and $K=20$ for all the results shown in this article. \par
After completing a successful training to start the prediction process, we take any random point from the available trajectory as the initial condition and warm-up the machine for first $w_a=500$ time steps replacing the new parameter value. Particularly, we choose the initial condition as $(4,0,0,0.1)$. Then we allow machine to generate the predicted trajectory at this new parameter in an auto iterative way following eq.~\ref{eq:prediction2}-\ref{eq:prediction4}. Here, in fig.~\ref{fig:predicted_dynamics}, we show the dynamics of the training data $(\epsilon=0.4,0.38,0.36)$ in the first row, and try to predict the dynamics at diverse values of $\epsilon$. Figs.~\ref{fig:predicted_dynamics}(d), (e), (f) depict the original system dynamics and the corresponding predicted dynamics are plotted in Figs.~\ref{fig:predicted_dynamics}(g), (h) and (i) respectively. We observe that, machine can capture diverse dynamics extremely well which are not at all revealed to the machine during training. Machine is only aware of the periodic dynamics with different amplitude and corresponding system parameters. Any other information is not disclosed from the earlier. Interestingly, depending only on the new system parameter value, machine is able to predict more complex dynamics like chaos (see Fig.\ref{fig:predicted_dynamics}(g)) as well as simple dynamics like steady state (see Fig.\ref{fig:predicted_dynamics}(i)) extremely well even at distant parameters from the training data. 
\subsection{Prediction of bifurcation diagram and entropy}
\begin{figure}
    \centering
    \includegraphics[width=0.5\textwidth]{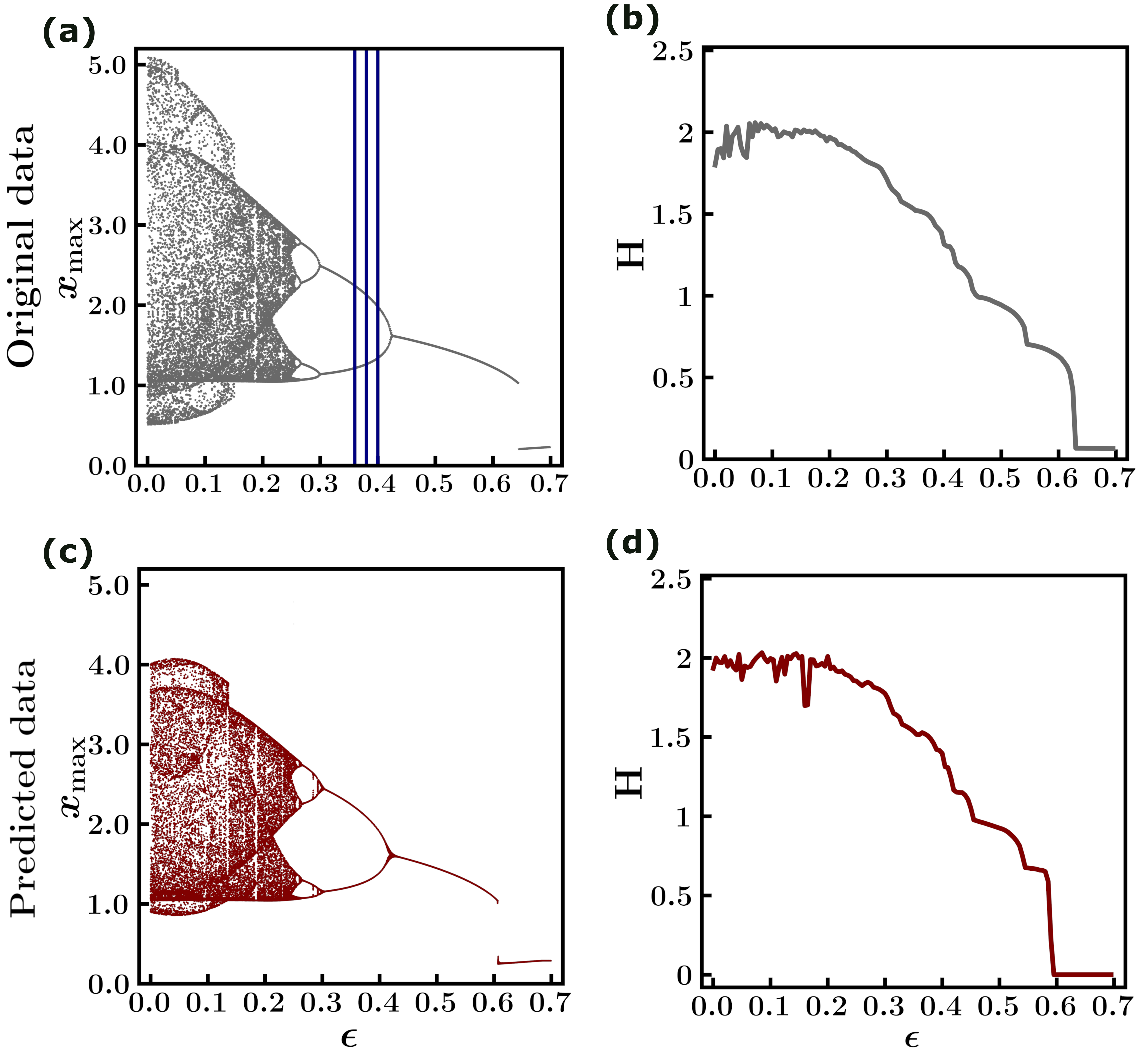}
    \caption{(a) Original bifurcation diagram and (b) entropy of the original system~\eqref{eq:original_model1} are plotted with respect to $\epsilon$. Three blue vertical lines in (a) represent the parameter values of the training dynamics. (c) Predicted bifurcation diagram and (d) predicted entropy are obtained from the ESN predicted time series.}
    \label{fig:bifurcation}
\end{figure}
We extend this study for the whole range of parameter space.  We follow the same training scheme discussed in the previous section for predicting the dynamics of the system for a large set of values of $\epsilon$. In Fig.\ref{fig:bifurcation}(a), we plot the original bifurcation diagram and in Fig.\ref{fig:bifurcation}(c), we plot the bifurcation diagram obtained from the ESN predicted dynamics. Here, we train the machine only with the dynamics at $\epsilon=0.4, ~0.38, ~0.36$ (see Figs.\ref{fig:predicted_dynamics}(a), (b), (c)) and the blue vertical lines in Fig.\ref{fig:bifurcation}(a). Now, machine can easily predict the dynamics throughout the whole parameter space. Here, machine is not only able to predict the dynamics when the parameter is quite far from the training parameter, interestingly, it can also capture the bifurcation scenario of period doubling route towards chaos. The transition points are captured almost perfectly. Therefore, we reproduce the whole bifurcation diagram of the system with simple ESN architecture and only a small number of periodic training dynamics . \par
Furthermore, We calculate the entropy profile with respect to $\epsilon$ from the original and predicted time series data as depicted in Fig.\ref{fig:bifurcation}(b) and Fig.\ref{fig:bifurcation}(d). The entropy for a given trajectory is defined\cite{shannon1948mathematical} as
\begin{equation}
    H = - \sum_{i=1}^n p_i~log(p_i),
\end{equation}
where, $p_i$ is the nonzero probability of $i^{th}$ state of the phase space being occupied when the whole state space volume is discretized by sufficiently small unit volumes. $n$ is the total number of states with nonzero probability. We observe that the entropy profile of the predicted dynamics matches with that of the original system very well, which verifies the degree of phase space occupancy of the predicted trajectory for the whole range of the parameter $\epsilon$.
\subsection{Prediction of basin for unknown attractors}
\begin{figure}
    \centering
    \includegraphics[width=0.5\textwidth]{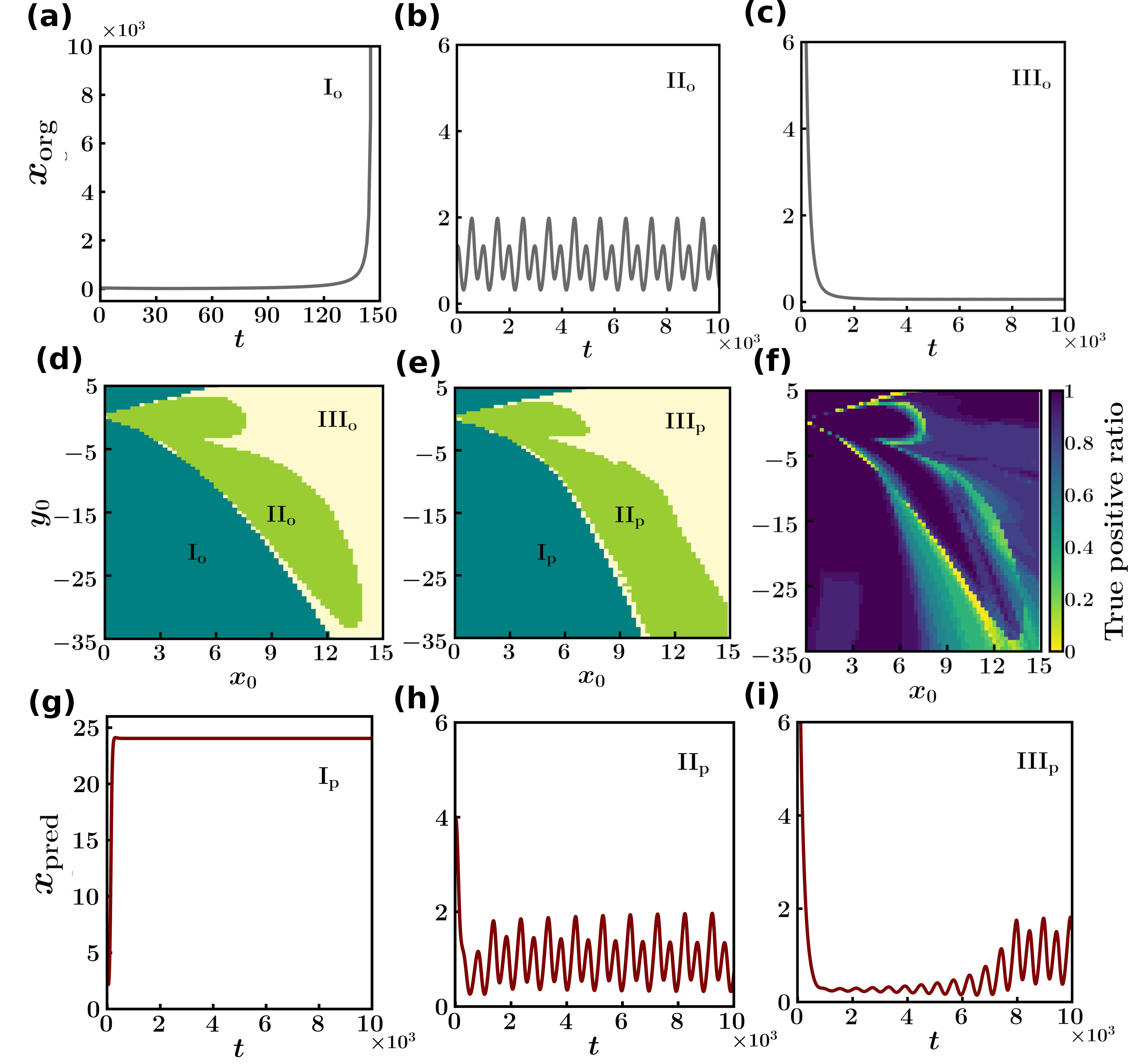}
    \caption{Basin prediction for system~\eqref{eq:original_model1}. (a), (b), (c) Three original dynamics are presented for three different region of the (d) original basin. Region $\rm{I_o}$, region $\rm{II_o}$ and region $\rm{III_o}$ correspond to (a) unbounded solution, (b) oscillation and (c) steady state. (e) Predicted basin and the corresponding predicted dynamics for region $\rm{I_p}$, region $\rm{II_p}$, region $\rm{III_p}$ are shown in (g), (h), (i). True positive ratio for the basin prediction is depicted in (f).  }
    \label{fig:basin}
\end{figure}
We continue this study with the prediction of basin for different attractors using the same architecture of the reservoir. Here, the system~\eqref{eq:original_model1} have a multistable region in the $(\beta-\epsilon)$ parameter space as shown in Fig.\ref{fig:original_model1}(a), highlighted with the yellow color. We randomly choose a parameter set $(\beta,\epsilon)=(0.15,0.4)$ from the bistable region for identifying basins. In Fig.\ref{fig:basin}(d), we plot basin of the original system. Three different colored basins exhibit three different dynamics (time series are presented in Figs.\ref{fig:basin}(a), (b), (c)). For the initial condition in region $\rm{I_o}$ (shown by teal color), system blows up towards infinity (see Fig.\ref{fig:basin}(a)). Region $\rm{II_o}$ (highlighted with green color) produces oscillating dynamics which is depicted in Fig.\ref{fig:basin}(b). Lastly, region $\rm{III_o}$ gives a stable steady state dynamics as shown in Fig.\ref{fig:basin}(c).\par
To start the prediction of basin through ESN, one should notice that, the chosen training parameter set is not very far from the parameter at which the basins are to be identified. Considering the complexity of multistable system, the basin prediction task at a completely unknown parameter is a quite challenging task itself. However, one can always increase this parameter drift by sacrificing the accuracy and using larger training data set with greater computational cost. Only one type of dynamics (feasible to choose the oscillatory dynamics) we use for training and no other information of the system is revealed to machine. Here, we choose the oscillatory dynamics at the parameter values $\epsilon=0.43,~ 0.42,~ 0.41$ as the training data and want to predict the basins for all the three different dynamics at $\epsilon=0.4$. We follow the same procedure for training as previously mentioned and after a successful training, we start predicting the dynamics at $\epsilon=0.4$. Here, we fix the machine hyper-parameters at $\rho=0.06, ~\sigma=0.5057,~a=0.385,~b=5\times 10^{-8}$. 
Now, to identify the basins we start the calculations from the space initial conditions $(x_0,y_0)\in [0,15]\times[-35,5]$. Interestingly, we find three different predicated dynamics depending only on the initial conditions. First one is the unusual dynamics like Fig.\ref{fig:basin}(g), which is not bounded (approaches near $x=25$) into the usual range of the system trajectories. Second one is the oscillation (as shown in Fig.\ref{fig:basin}(h)). Here, we need to focus on the transient dynamics, because in Fig.\ref{fig:basin}(h), the oscillation starts immediately after starting the prediction. On the other hand, in Fig.\ref{fig:basin}(i), we find the trajectory initially generates negligible oscillations (almost steady state) for large time points and after that it produces oscillations. For this study, we take $2500$ time points as the threshold for classifying two different dynamics. Producing subthreshold oscillation (we choose $x<1.4$) for initial 2500 time points fells it into the third type of dynamics (\ref{fig:basin}(i)). Based on these three types of predicted dynamics, we obtain Fig.\ref{fig:basin}(e), the predicted basin. We find the machine is able to identify the basins for unbounded dynamics very well (Region $\rm{I_p}$ in Fig.\ref{fig:basin}(e)). Region $\rm{II_p}$ in predicted basin, ESN generated dynamics is same as the original system dynamics. Lastly, for region $\rm{III_p}$ exhibits dynamics as shown in Fig.\ref{fig:basin}(i). Here, machine is not generating a stable steady state dynamics, but it initially produces transient small oscillations, depending on that we can separate region $\rm{III_p}$ in Fig.\ref{fig:basin}(e).\par
Furthermore, in the process of generating reservoir some randomness enter into the system. To overcome the impact of randomness over the obtained results and to check the robustness of our scheme, we repeat the same investigation for a sufficient times (say $\rm{N}$) for independently generated reservoir keeping the ESN hyperparameter fixed. Depending on number of correct prediction (say $\rm{M}$), we calculate the quantity, $\rm{True~positive~ratio} = \frac{M}{N}$. 
If the ratio goes to unity, it means every time machine is able to produce the right prediction and value of this ratio is zero means machine is not able to predict the dynamics correctly for a single time. Intermediate values of the ratio shows the level of producing the correct prediction by ESN depending on the intrinsic randomness. Here, we take N=100. In Fig.\ref{fig:basin}(f), we plot the true positive ratio at every initial conditions for the whole space on a 0 (yellow) to 1 (blue) scale. Machine is able to predict the whole space of the basin quite perfectly. Only near boundaries when transition occurs from one dynamics to other, the accuracy level of the predicted dynamics is at lower side. Therefore, we can conclude that excluding the boundary regions, over the space of whole initial conditions basins for different dynamics are considerably well predicted.  

\subsection{Another bistable system}\label{sec:second_model}
\begin{figure}
    \centering
    \includegraphics[width=0.5\textwidth]{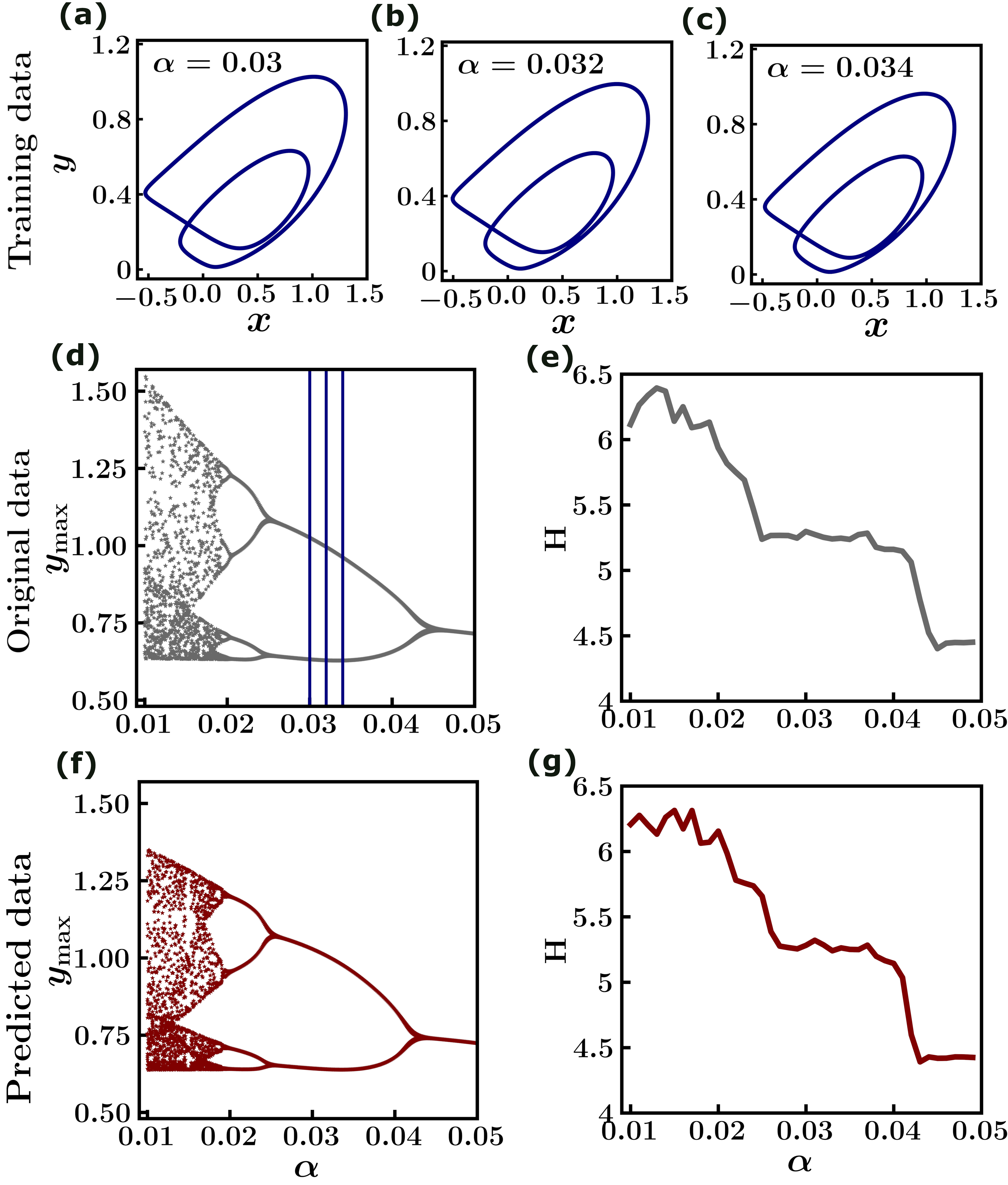}
    \caption{Prediction of the bifurcation diagram and entropy. (a), (b), (c) Oscillatory dynamics for training are shown on $x-y$ phase space. (d) Original bifurcation diagram and (e) original entropy are plotted with respect to the parameter $\alpha$. Three blues lines in (a) represent the parameter values of the training dynamics. Corresponding predicted bifurcation diagram and the entropy are plotted in (f) and (g) respectively. }
    \label{fig:bifurcation2}
\end{figure}
We generalize the results by considering another system~\cite{wang2012chaotic} having bistability as follows:
\begin{equation}\label{eq:original_model2}
\begin{aligned}
\dot{x} &=yz+\alpha\\
\dot{y} &=x^2-y\\
\dot{z} &=1-4x\\
\end{aligned}
\end{equation}
where, for $\alpha<0$ the system possesses an unstable fixed point, but for $\alpha>0$ the system enters into a bistable regime, where a stable fixed point $(\frac{1}{4},\frac{1}{16},-16\alpha)$ and a stable oscillator coexist.\par
Similarly for this model, we train the machine with a few simple dynamics of the system with the parameter values of $\alpha$ and let the machine predict the dynamics for whole range of $\alpha$. Here, we take the periodic dynamics (see Figs.\ref{fig:bifurcation2}(a), (b), (c)) for $\alpha=0.03,~0.032,~ 0.034$ to train ESN, which are also shown by the three blue vertical lines in Fig.\ref{fig:bifurcation2}(d). The corresponding hyperparameters are as follows: $\rho=0.5057,~ \sigma=0.35 ,~ a=0.25 ,~ b=5\times10^{-6}$. After completing the training procedure, we change the new parameter values to the machine input, and continuing the same process for the whole parameter range, we obtain the predicted bifurcation diagram depending only on the values of $\alpha$. In Fig.\ref{fig:bifurcation2}(d), we plot the original bifurcation diagram and Fig.\ref{fig:bifurcation2}(f) is the corresponding machine predicted diagram. We observe that machine identifies the transition towards higher periodic order quite accurately. Moreover, we plot the corresponding entropy of the original and the machine predicted data in Fig.\ref{fig:bifurcation2}(e) and in Fig.\ref{fig:bifurcation2}(g) respectively. The original and machine predicted entropy follow almost similar path with respect to $\alpha$.  \par  
\begin{figure}
    \centering
    \includegraphics[width=0.5\textwidth]{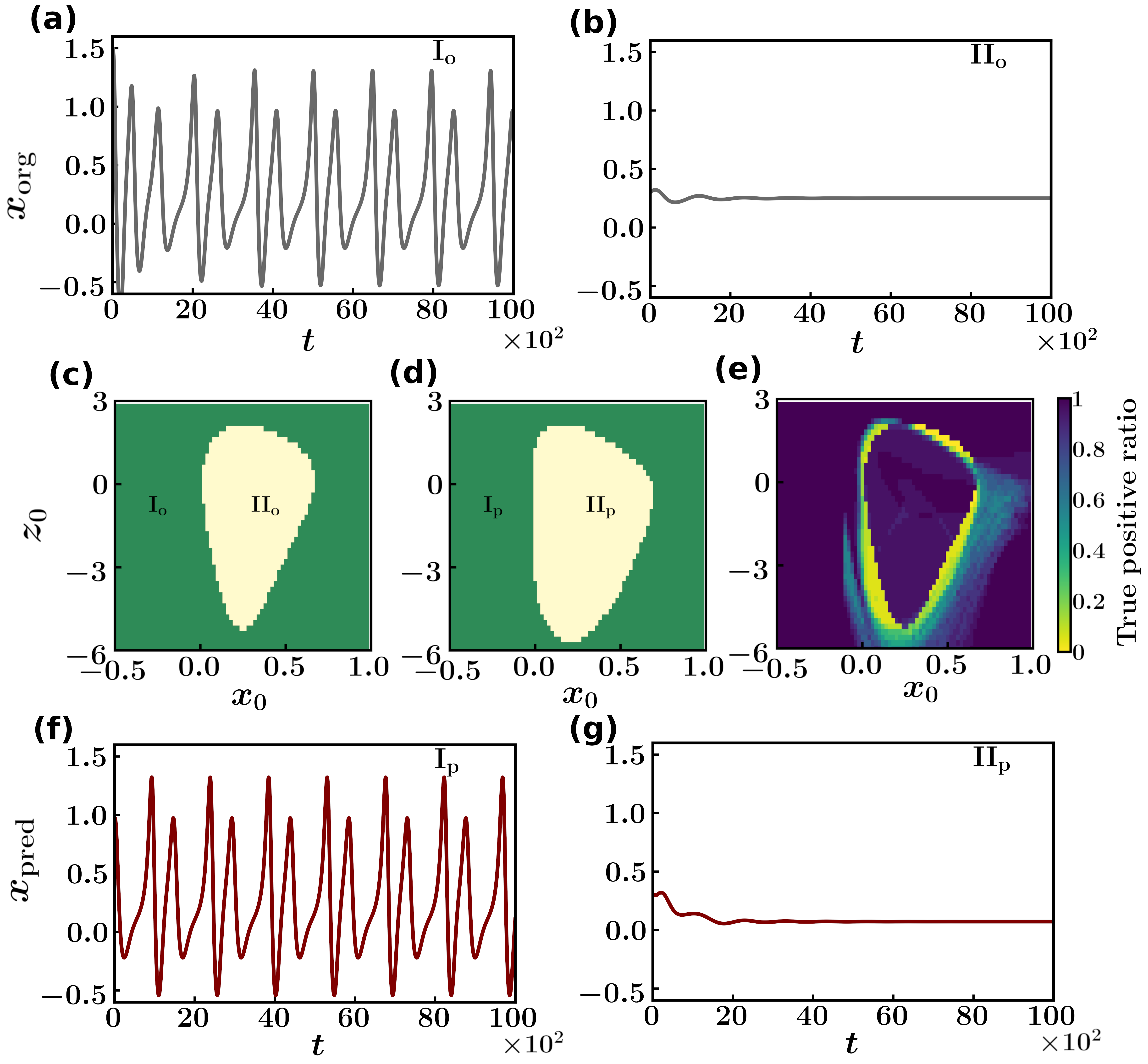}
    \caption{Basin prediction for system~\eqref{eq:original_model2}. Time series of the original dynamics (a) Oscillation and (b) fixed point dynamics corresponding to two different region of the (c) original basin. (d) Predicted basin with the corresponding predicted dynamics (f) oscillation and (g) steady state. (e) True positive ratio for the prediction of the basin.  }
    \label{fig:basin2}
\end{figure}
The system possesses a bistability between two stable dynamics: (i) stable oscillation (see Fig.\ref{fig:basin2}(a)) and (ii) stable steady state (see Fig.\ref{fig:basin2}(b)). We plot the corresponding original basin in Fig.\ref{fig:basin2}(c) over the initial conditions $(x_0,z_0)\in [-0.5,1.0]\times[-6,3]$   , where region $\rm{I_o}$ resembles the initial conditions for oscillating dynamics and region $\rm{II_o}$ is the basin for steady state. The original basin is corresponding to $\alpha=0.03$. For training we take the oscillation dynamics at $\alpha=0.033,~ 0.032,~0.031$. In this case we set the machine hyperparameters as: $\rho=0.5057,~ \sigma=0.06,~ a=0.385 ,~ b=9\times 10^{-7}$. Now ESN generates two types of dynamics on the space of initial conditions. One oscillatory dynamics (shown in Fig.\ref{fig:basin2}(f) and one stable steady state (shown in Fig.\ref{fig:basin2}(g). Now, we separate the basins based on these two dynamics, and finally we obtain the predicted basin for this system in Fig.\ref{fig:basin2}(d). Now, to verify the robustness under the system's intrinsic randomness, we plot the true positive ratio for the basin in Fig.\ref{fig:basin2}(e) and similar to the previous system  we find that near the boundary true positive ratio remains low. However, excluding the boundary region machine is able to capture two separate dynamics considerably well depending only on the initial conditions. 


\section{conclusion}\label{sec:conclusion}
In this study, we investigated a strategy to identify diverse dynamics in a system through a parameter-aware ESN approach. Using this technique, learning from a few training dynamics ESN can build the dynamics for any parameter value even if it lies considerably far from the training parameter set. We elaborately explained the whole training and prediction process for obtaining the diverse predicted dynamics. Without being exposed to any other additional information, machine is able to train itself in such a way that it can capture the nature of complex dynamics for a broad range of parameter values. Exploiting this scheme, we are able to draw the predicted bifurcation diagram which almost follows the original dynamics. Another significant direction of this study is to identify the basins for multistable attractors. It is well known that there are several systems appearing in many areas possess bistability or multistability for a certain parameter space. Also, without any prior knowledge of the basin these systems can produce various undesired situations in practice. In this context, we gave a recipe for locating basins of multiple co-existing attractors using the dynamical information of only one of them. Training with only one time series at few parameter values each is enough for the machine to be trained such that it can predict the dependency of the system dynamics on the initial conditions. Machine is able to generate different dynamics if we start the prediction process from different initial conditions and eventually the basin is also predicted without putting any extra effort. In this context, a very recent study\cite{gauthier2022learning} has attempted to show the utility of a different machine learning scheme for locating an unseen attractors of a multistable system. Thus, machine learning approach for predicting multistability contributes greatly in better understanding of complex dynamical systems. 
\section*{Conflict of Interest}
The authors have no conflict to disclose.

\section*{Author Contributions}
M.R: Conceptualization (equal); Methodology (equal); Formal analysis (lead); Investigation (lead); Software (lead); Visualization (lead); Validation (equal); Writing – original draft (lead). 
S. M.: Conceptualization (equal); Methodology (equal); Formal analysis (lead); Investigation (lead); Software (lead); Visualization (lead); Validation (equal); Writing – original draft (lead).  
C.H.:  Conceptualization (equal); Methodology (equal); Validation (equal); Writing – review and editing (equal). 
A.P.: Conceptualization (equal); Supervision (equal); Writing – review and editing (equal).
N.V.K.: Supervision (equal); Writing – review and editing (equal).
M.D.S.: Conceptualization (equal); Methodology (equal); Validation (equal); Supervision (equal); Writing – review and editing (equal).

\section*{Data Availability}
The data that support the findings of this study are available
from the corresponding author upon reasonable request.

\section*{Acknowledgement}
M.R. and M.D.S. are financially supported by Department of Science and Technology (DST), India under the Indo-Russian Joint Research Programme (No. INT/RUS/RSF/P-18). S.M. and M.D.S. acknowledge SERB, Department of Science and Technology (DST), India (DST - SERB (CRG/2021/003301)). C.H. is supported by DST-INSPIRE-Faculty grant (Grant No. IFA17-PH193).

\section*{references}
\bibliography{aipsamp}

\end{document}